\documentclass[a4paper,11pt]{article}
\pdfoutput=1 

\usepackage{jcappub} 

\usepackage[T1]{fontenc} 

\title{\boldmath Searching for  Fast Neutrino Flavor Conversion Modes in Core-collapse Supernova Simulations}

\author[a,b]{Sajad Abbar}

\affiliation[a]{Max-Planck-Institut f\"ur Physik (Werner-Heisenberg-Institut),\\ F\"ohringer Ring 6, 80805 M\"unchen, Germany}
\affiliation[b]{Astro-Particule et Cosmologie (APC),\\ CNRS UMR 7164, Universit\'e Denis Diderot, 75205 Paris Cedex 13, France }

\emailAdd{abbar@mpp.mpg.de}

\abstract{Neutrinos propagating in  dense neutrino media such as 
 those in core-collapse supernovae
 can experience fast flavor conversions 
  on scales
much shorter than those expected in vacuum.
It is believed that  a necessary condition for the occurrence of 
fast  modes is that the angular distributions of $\nu_e$
and $\bar\nu_e$ cross each other. However, most of the state-of-the-art
supernova simulations do not provide such detailed angular information and instead,
consider only a few moments of neutrino angular distributions. 
We here propose
an efficient method 
 to use these available few moments to search for fast modes
in supernova simulations.
 Our  method, which is
based on searching for crossings in the angular distributions,
can work with any number of   moments provided by the simulation though
 a larger number of  crossings can be captured
 when higher moments are available. 
}

\begin{document}
\subheader{\hfill MPP-2020-11}
\maketitle
\flushbottom


\section{Introduction}

Core-collapse supernova (CCSN) explosions
are among the most
energetic astrophysical phenomena.  
The  explosion is caused by the death of a massive star
with a mass larger than $8-10\ \rm{M}_{\odot}$ which runs out of
fuel in its core~\cite{Colgate:1966ax, Bethe:1984ux, 
 Janka:2012wk, Burrows:2012ew}.
In this process, 
neutrino emission is a major effect  
and a huge
number of neutrinos ($\sim~10^{58}$) are emitted within the time 
interval of a few (tens of)
seconds. 

Because their vacuum and flavor eigenstates are not coincident,
neutrinos can experience flavor oscillations while propagating.
In particular, they can oscillate collectively  in the SN
environment, due to their coherent forward scatterings by the 
high density background neutrino 
gas~\cite{Pastor:2002we,duan:2006an, duan:2006jv, duan:2010bg,
  Chakraborty:2016yeg, Qian:1996xt}.
Collective neutrino  oscillations is a nonlinear phenomenon in which neutrinos
and antineutrinos with different momenta get coupled to each other.

Neutrino oscillations can change the spectra of the neutrinos and antineutrinos
and consequently, could remarkably affect the physics
of CCSNe. 
 Firstly, it can influence heavy elements nucleosynthesis
 by  modifying the neutron-to-proton ratio. Secondly, 
it could affect the SN dynamics by changing the neutrino energy deposition into
the shock wave. Specifically,  in the popular so-called delayed explosion mechanism  
 the SN explosion is aided by absorbing a fraction of the energy of neutrinos 
emitted from the SN core.
Finally, it can modify the potential SN neutrino signal 
which may be detected on earth.

Although the first studies on neutrino evolution in CCSNe
 were carried out in maximally symmetric models, e.g.
 the stationary spherically symmetric bulb 
 model~\cite{duan:2006an, duan:2006jv, duan:2007sh, dasgupta:2009mg, duan:2010bg,
 Galais:2011gh, Duan:2007bt, Galais:2011gh, Duan:2015cqa}, 
 it was then realized that 
 the spatial/temporal symmetries are not compatible with collective neutrino oscillations
 in the sense that   any symmetries imposed initially on the surface of the neutrino emitter
 can be broken spontaneously in a dense neutrino 
 gas~\cite{raffelt:2013rqa, duan:2013kba, duan:2014gfa,
 abbar:2015mca, Abbar:2015fwa, chakraborty:2015tfa, Chakraborty:2016yeg,
 Dasgupta:2015iia, Mirizzi:2015fva,Martin:2019gxb, Martin:2019kgi}.
 
Moreover, it has been  shown that neutrinos can also undergo fast flavor
conversions in dense neutrino 
media~\cite{Sawyer:2005jk, Sawyer:2015dsa,
 Chakraborty:2016lct, Izaguirre:2016gsx,  Wu:2017qpc,
  Capozzi:2017gqd, Richers:2019grc,  
  Dasgupta:2016dbv, Abbar:2017pkh, Abbar:2018beu, Capozzi:2018clo,
 Martin:2019gxb, Capozzi:2019lso, Doring:2019axc, Chakraborty:2019wxe, Johns:2019izj,
 Shalgar:2019qwg, Cherry:2019vkv}.  
Remarkably, fast modes occur on scales $\sim G_{\rm{F}}^{-1} n_{\nu}^{-1}$
with $n_\nu$ and $G_{\rm{F}}$ being the neutrino number density and the 
Fermi coupling constant, respectively. Such scales can be as short as a few
cm's just above the photo-neutron star (PNS).
This must be compared with
 the traditional collective modes which occur on scales 
 determined by the neutrino vacuum frequency, $\omega = \Delta m^2/2E$, 
which would be 
$\sim \mathcal{O}(1)$ km for a 10 MeV neutrino and atmospheric mass
splitting.

It is currently believed that a necessary condition (maybe even
sufficient  in  reasonable cases)  for the occurrence of fast modes
is that the angular distributions of $\nu_e$ and $\bar\nu_e$
cross each other (assuming $\nu_x$ and $\bar\nu_x$ have similar  
distributions)~\cite{Abbar:2017pkh, Capozzi:2019lso}. 
In other words, fast modes can exist provided that
 the angular distribution of the neutrino electron lepton 
number (ELN)  defined as
\cite{Izaguirre:2016gsx},
\begin{equation}
  G(\mu) =
  \sqrt2 G_{\mathrm{F}}
  \int_0^\infty \int_0^{2\pi} \frac{E_\nu^2 \mathrm{d} E_\nu \mathrm{d} \phi_\nu}{(2\pi)^3}
        [f_{\nu_e}(\mathbf{p})- f_{\bar\nu_e}(\mathbf{p})],
 \label{Eq:G}
\end{equation}
crosses zero at some $\mu=\cos\theta_\nu$\footnote{Note that we 
have integrated over $\phi_\nu$ here. Otherwise,
ELN crossings can exist in  $\phi_\nu$ as well.}. 
Here, $E_\nu$, $\theta_\nu$ and $\phi_\nu$ are the neutrino energy, and 
the zenith and azimuthal angles of the neutrino velocity, respectively, 
and  $f_{\nu}$'s are the neutrino 
occupation numbers where
 we have assumed $f_{\nu_x}(\mathbf{p})= f_{\bar\nu_x}(\mathbf{p})$.

Recently, several groups have reported the occurrence of fast modes  
in SN simulations~\cite{Abbar:2018shq, Abbar:2019zoq, 
 DelfanAzari:2019tez, Nagakura:2019sig, 
Morinaga:2019wsv,   Glas:2019ijo, Rampp:2002bq}.
Interestingly, it turns out that  fast
modes can exist in three different regions in the SN environment, namely 
within the neutrino decoupling region~\cite{Abbar:2018shq, Abbar:2019zoq, 
 Nagakura:2019sig}, 
inside the PNS~\cite{ Abbar:2019zoq, 
 DelfanAzari:2019tez, Glas:2019ijo},  and
in the pre-shock SN region~\cite{Morinaga:2019wsv}. 
Nevertheless, the potential physical implications 
of the presence of fast modes inside the PNS and
in the pre-shock SN region 
 is  currently  not clear.
On the one hand, the ELN crossings inside the PNS
occur at the SN zones where the neutrino gas is non-degenerate.
This   prevents any significant flavor conversions 
since all flavors have  almost identical distributions. In addition,
such a high degree of non-degeneracy  allows for the existence of 
rapid conversion modes even without the presence of ELN crossings, 
as discussed in Ref.~\cite{Abbar:2019zoq}.
Hence, inside the PNS, 
fast modes is not the only phenomenon that can lead
to  flavor conversions on short scales. 
On the other hand, the ELN crossings in the pre-shock region seem
to be extremely  narrow and consequently, any fast conversion rates
therein  can  at most be comparable to the ones of slow 
modes (see Fig. 2 of Ref.~\cite{Morinaga:2019wsv}).

Despite the important observation that fast modes can exist in the SN environment,
all such studies are  limited  by one or more
of the following factors:
being performed in one or two dimensions (Refs.~\cite{
 DelfanAzari:2019tez, Nagakura:2019sig, 
Morinaga:2019wsv}), capturing only fast modes
 inside the PNS (Refs.~\cite{
 DelfanAzari:2019tez, Glas:2019ijo})
or being based on the post-processing calculations 
(Refs.~\cite{Abbar:2018shq, Abbar:2019zoq}).
In particular,  the multidimensional (multi-D)   
CCSN simulations  providing  full neutrino angular 
distributions, which have just become accessible, are  
confined to the 2D models and also have relatively low angular resolutions.
Thus, it is safe to say that our understanding of the characteristics of fast modes
in   realistic  SN models  is still very limited.

In addition, such detailed angular information is
not available in most of the state-of-the-art 3D CCSN simulations
due to the unbearable computational 
costs~\cite{Tamborra:2014aua, Bruenn:2014qea, OConnor:2015rwy,
Richers:2017awc, Vartanyan:2018xcd,  Just:2018djz, Pan:2018vkx}. 
Instead, the
neutrino transport is treated  by considering 
a few number of angular moments
of the neutrino phase-space distributions. 
One can, indeed, express the Boltzmann equation
in terms of an infinite series of equations for neutrino angular moments
in which, the evolution equation of each moment is only coupled 
to a few of its neighboring moments.
In order to reduce
the computational cost of solving the Boltzmann  equation, one
can then employ some algebraic closure methods to close
the equations for the first few moments, 
 by assuming some analytical forms for some of the higher moments.

This means that in the moments method, instead of having ELN distribution,
one has only a few of its angular moments defined as,
 \begin{equation}
I_n = \int_{-1}^{1} \mathrm{d}\mu\ \mu^n\ G(\mu).
\end{equation}
For instance, in the $M_1$ closure scheme, $I_0$, $I_1$, $I_2$ and
$I_3$ are considered in the neutrino transport, out of which $I_2$ and
$I_3$ are  related analytically to $I_0$ and $I_1$~\cite{Just:2015fda, Murchikova:2017zsy}.
It is of great importance  to note that in the moments method, it is \emph{not} assumed that
higher moments are zero or can be ignored. They are, in principle, nonzero
and evolve in time since the information can flow from the equations
of lower moments to the ones of higher moments, but 
this information is just not provided in the simulation.

Although a huge part of the angular information is lost by considering
only a few angular moments, one can still use the limited available 
information to assess the possibility of the occurrence of fast modes.
This is  of utmost importance because  
the current state-of-the-art 3D CCSN simulations are based on the moments method.
It is, indeed, possible to express the criteria 
for the instability of some of the flavor conversion modes
in terms of a few neutrino angular moments~\cite{Dasgupta:2018ulw, Johns:2019izj}. 
Specifically, 
 in Ref.~\cite{Glas:2019ijo}, the authors indicated that   a number of ELN 
crossings inside the PNS can be captured by using this method. 

In this study, we propose a new method to search for neutrino fast flavor conversion 
modes by analyzing  a few neutrino angular moments. 
 Unlike the methods proposed in Refs.~\cite{Dasgupta:2018ulw, Johns:2019izj}
 which are based on 
 the instability of a few specific modes, 
our method focuses on looking for  crossings in the ELN angular 
distribution (see Sec.~\ref{sec:ELN}).
We show that our method is comparatively sensitive to
narrow ELN crossings  above the neutrinosphere 
and within the neutrino decoupling region,
in contrast to the one  based on 
 the instability of the so-called zeroth mode~\cite{Dasgupta:2018ulw}
 which is mostly sensitive to the widest
 ELN crossings anticipated inside the PNS.
This makes our method significantly stronger 
 in capturing the ELN crossings and the consequent fast modes in the
SN environment.

\section{Searching for  ELN Crossings} \label{sec:ELN}
Our method is based on the following simple observation:
\\

\textbf{Theorem.} Let's assume that 
  the ELN angular distribution, $G(\mu)$, does not have any crossings. Then $I_0$ and $I_\mathcal{F}$ defined as,
 \begin{equation}\label{eq:I}
I_\mathcal{F} = \int_{-1}^{1} \mathrm{d}\mu\ \mathcal{F}(\mu)\  G(\mu),
\end{equation}
must have the same sign for {\it{any}} function $\mathcal{F}(\mu)$
which is positive in the interval $[-1,+1]$. 

 One  can  \emph{equivalently} say that if there exists a positive  function, $\mathcal{F}(\mu)$,
 for which $I_0 \cdot I_\mathcal{F} < 0$, then $G(\mu)$ 
  definitely features a crossing in  the interval $[-1,+1]$.
 \\

\textit{Proof.} The theorem is almost trivial. If $G(\mu)$ is always positive in the interval $[-1,+1]$,
then  $I_0$ and $I_\mathcal{F}$ 
are both positive for any  positive function  $\mathcal{F}(\mu)$. 
A similar argumnat applies when  $G(\mu)$ is always negative
 in the interval $[-1,+1]$.
  \\

In order to take  advantage of the information provided in CCSN simulations,
we  choose $\mathcal{F}(\mu)$ to have the form~\footnote{Or equivalently
the Legendre polynomials, $P_n(\mu)$'s, can be used
\begin{equation}\label{eq:F}
\mathcal{F}(\mu) = \sum_{n=0}^{n=N} a_n P_n(\mu).
\end{equation}
Note, however, that such an expansion
can not be used to reconstruct an approximate angular distribution
of ELN. This simply arises from the fact that the higher polynomials
are not necessarily negligible (they are not just provided by the simulation).}
 \begin{equation}\label{eq:F}
\mathcal{F}(\mu) = \sum_{n=0}^{n=N} a_n \mu^n,
\end{equation}
where $a_n$'s are some arbitrary coefficients (for which $\mathcal{F}(\mu)$ is positive)
and  $N$ can be any number from 1 to the maximum number of 
angular moments available in the simulation.
Inserting  this $\mathcal{F}(\mu)$  into Eq.~(\ref{eq:I}) results in
 \begin{equation}
I_\mathcal{F} = \sum_{n=0}^{n=N} a_n I_{n},
\end{equation} 
so that $I_\mathcal{F}$ can be obtained in terms of the neutrino
angular moments which are provided by the CCSN simulation.
 Any sign difference  between $I_0$ and   $I_\mathcal{F}$
 for some positive $\mathcal{F}(\mu)$
in a SN zone would be a definite sign of 
  the existence of ELN crossing at that zone.
  
For example, if only  $I_0$ and $I_1$ are considered in the SN simulation,
then $\mathcal{F}(\mu) = a_0 +a_1 \mu$, for any $a_0$
and $a_1$ for which $\mathcal{F}(\mu)$ is positive in the interval $[-1,+1]$.
For instance, one can choose $\mathcal{F}(\mu) = 1 + \mu$ or 
$\mathcal{F}(\mu) =1 - \mu$ 
 for which 
$I_\mathcal{F} = I_0 +I_1$ and  $I_\mathcal{F} =I_0-I_1$, respectively. 
If $I_2$ is also provided, apart from $I_\mathcal{F} $'s mentioned above,
one can also consider 
 any combinations of $a_0$, $a_1$ and $a_2$ for which 
 $\mathcal{F}(\mu)  = a_0 + a_1 \mu + a_2 \mu^2$
is  positive in the interval $[-1,+1]$~\footnote{Note that it
  is always possible
to decrease the number of degrees of freedom by taking out 
the (positive part of)  $a_N$ and fixing $a_0$. For 
instance, in the case previously mentioned where  
only  $I_0$ and $I_1$ are available, $\mathcal{F} = 1 + \mu$ and  
$\mathcal{F} = 1 - \mu$ are the only choices one can consider.}. 
 Note that the inequality proposed in 
 Ref.~\cite{Dasgupta:2018ulw}~\footnote{Although the criteria proposed in 
 Ref.~\cite{Johns:2019izj} may provide some useful insight in understanding
 the behavior of the homogenous mode, they are not, in general, helpful
 to capture fast modes, due to the approximations made in the analysis.},
\begin{equation}\label{eq:ineq}
(I_0 + I_2)^2 - 4I_1^2 < 0,
\end{equation}
which is based on the instability of the zeroth mode, 
$K=0$~\footnote{This zeroth mode is not the same as
the true spatially homogenous mode of the neutrino 
gas.},
 only corresponds to  two specific single points 
  in the 
parameter space of $a_0$, $a_1$ and  $a_2$
 with $a_0 = 1$, $a_1 = \pm 2$ and $a_2 = 1$,
  which corresponds to~\footnote{One may also need
  to consider the case $a_1 = a_0 = 0$ in proving
  the equivalence of Eqs.~(\ref{eq:ineq}) and (\ref{eq:ineq2}).}
\begin{equation}\label{eq:ineq2}
I_{\mathcal{F}} = I_2 \pm 2 I_1 +   I_0.
\end{equation}
In other words, our method offers an infinite number of
inequalities similar to Eq.~(\ref{eq:ineq}) to be checked for 
the occurrence of fast modes.
Note, however, that the methods based on the instability
of the flavor conversion modes can advantageously assess
the possibility of the occurrence of fast modes 
rather than focusing on the ELN crossings.
If higher moments such as $I_3$ are also available, 
one can follow the same argument and choose 
$\mathcal{F}(\mu)  = a_0 +a_1 \mu + a_2 \mu^2 + a_3 \mu^3$
 for any combinations of $a_i$'s for which $\mathcal{F}(\mu)$
   is  positive in the interval $[-1,+1]$.

\begin{figure}[tb!] 
\centering
\begin{center}
\includegraphics*[width=.5\textwidth, trim= 5 10 10 10,clip]{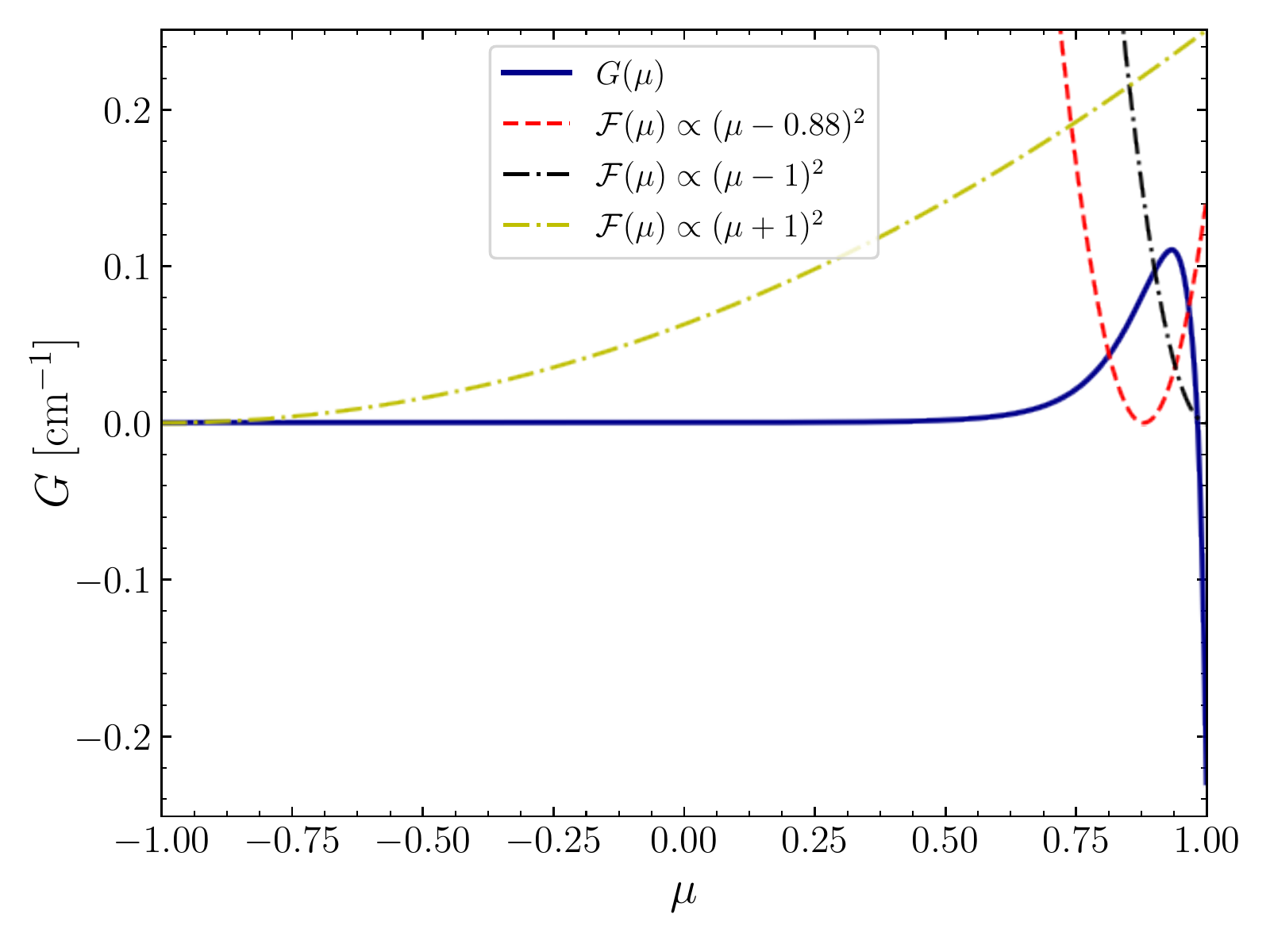}
\end{center}
\caption{
An example of an  ELN angular distribution with crossing (blue solid curve) obtained from 
Ref.~\cite{Abbar:2018shq},
and three choices of $\mathcal{F}(\mu)$, namely 
$\mathcal{F}(\mu) \propto (\mu-0.88)^2$ (red dashed line), and  
$\mathcal{F}(\mu) \propto (\mu \pm1)^2$
(black and yellow dash-dotted lines) corresponding to the 
instability condition of the zeroth mode in Eq.~(\ref{eq:ineq}).
Although the dash-dotted lines are not  appropriate choices for 
capturing the narrow ELN crossing and they miss it, the red curve can capture
the crossing  due to its suitable shape, as it gets larger for 
$\mu \gtrsim \mu_{\rm{crossing}}$ and has a minimum for $\mu \lesssim \mu_{\rm{crossing}}$
 where $G(\mu) $ has significant values.  This shows not all $\mathcal{F}(\mu) $'s are
 appropriate to capture ELN crossings in the SN environment.
}
\label{fig:G}
\end{figure}

Although one can in principle try any positive function $\mathcal{F}(\mu)$
in Eq.~(\ref{eq:I}),
not all $\mathcal{F}(\mu)$'s are suitable for capturing the potential
 crossings in the ELN distribution.
 For example for the case presented in Fig.\ref{fig:G},
 in spite of the fact that $\mathcal{F} = (\mu-0.88)^2$ can capture
 the crossing in the ELN distribution, the functions corresponding  
to the inequality in Eq.~(\ref{eq:ineq}) miss it.
The reason for this 
can be understood as follows. 
In order to have different
signs between $I_0$ and $I_\mathcal{F}$, $\mathcal{F}(\mu)$ must
be significant for the crossings region (here $G < 0$)  but 
 small anywhere else where $G(\mu)$
has significant (here) positive values. 
This way, the (here) positive peak gets significantly suppressed and also,
the crossing region can have a larger 
contribution to $I_\mathcal{F}$. This obviously leads to 
 a higher chance for having a sign difference between $I_0$ and $I_\mathcal{F}$.

As a matter of fact,
 the functions corresponding  
to the inequality in Eq.~(\ref{eq:ineq}) are  not, in general,  
the most suitable choices 
for capturing relatively narrow ELN crossings. 
Nevertheless,  they may be  appropriate 
 for capturing the ELN crossings
 inside the PNS because such crossings are not necessarily 
 narrow (see Fig.~8 of Ref.~\cite{Abbar:2019zoq}). One may equivalently say that
 the zeroth mode can be unstable only when the ELN crossing is
 wide enough (at
 least for $n_{\bar\nu_e}/n_{\nu_e}<1$)
 and in general, it may not be a good
 measure of  the overall flavor evolution of the neutrino gas within the
 decoupling region. 
 
 Not only does this Figure  clearly show the importance of choosing the right function $\mathcal{F}(\mu)$
 to be applied in Eq.~(\ref{eq:I}), but also  it provides  useful  insight on 
 how to choose the appropriate function.
 Specifically, once $I_3$ is also available, as in the $M_1$ closure, 
 choosing the suitable $\mathcal{F}(\mu)$ 
 could be a little tricky because there are more than one free parameters
 that must be checked. However, considering the intuition developed  here,
 the most appropriate functions  to be chosen are most likely the ones with minima 
 at $\mu$ close to one,
 to have the maximum suppression for the non-negligible positive part of 
 $G(\mu)$ (which is for $\mu \gtrsim 0.8$ in Fig.\ref{fig:G}).

\begin{figure*}[tb!] 
\centering
\begin{center}
\includegraphics*[width=1.\textwidth, trim= 10 10 10 10,clip]{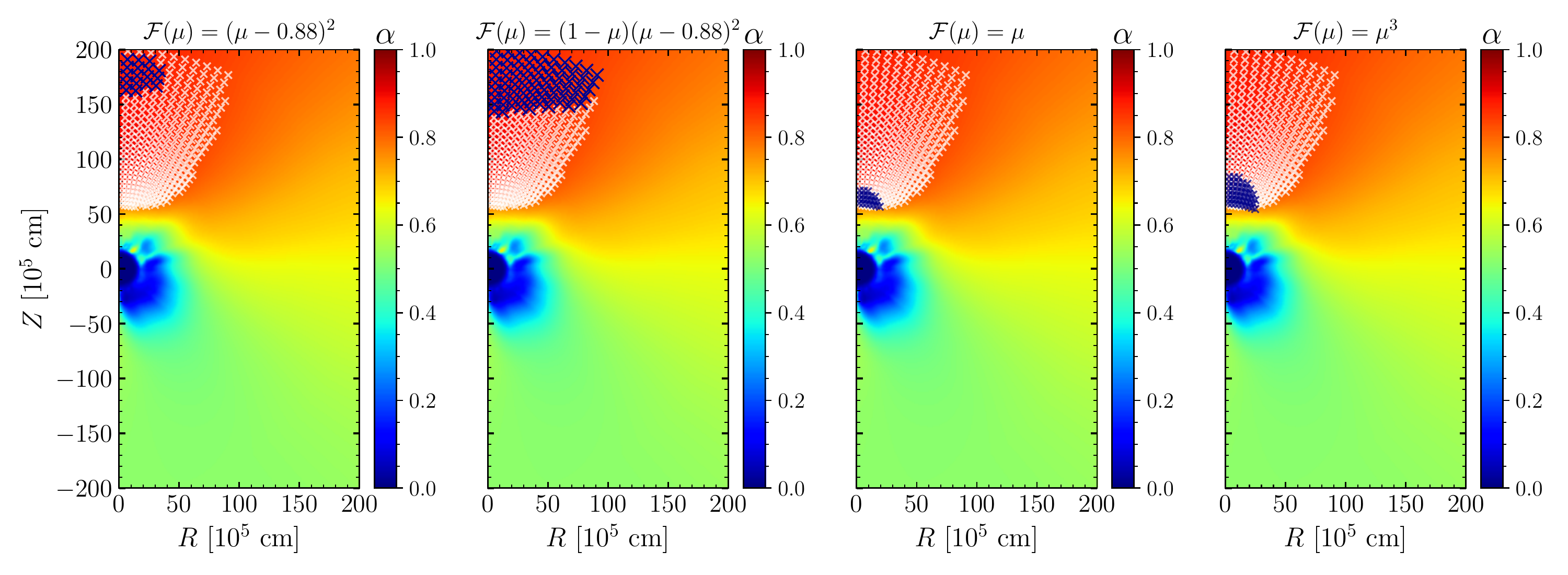}
\end{center}
\caption{
The $\nu_e-\bar{\nu}_e$ asymmetry parameter $\alpha$, defined as
$\alpha = n_{\bar\nu_e}/n_{\nu_e}$, in the $t_{\rm{pb}}=200$ ms snapshot
of the 2D SN model of an $11.2\mathrm{M}_{\odot}$ progenitor model~\cite{Abbar:2018shq}.
White Crosses indicate the ELN crossings found in the data and the blue crosses are
the ones which were captured by employing $\mathcal{F}(\mu) = (\mu-0.88)^2$,
$-(\mu-1)^3$, $\mu$ and $\mu^3$, respectively. 
At this time, 
neutrinos decouple from matter at radius $\sim 50-70$ km
depending on their flavors and energies.}
\label{fig:2D}
\end{figure*}

 To demonstrate the  practicability  of our method, we tested it for 
 some SN realistic neutrino angular distributions  studied in 
 Refs.~\cite{Abbar:2018shq, Abbar:2019zoq}.
These distributions were obtained by solving the Boltzmann transport equation 
(without any flavor transformation) for the fixed supernova profiles 
taken from the representative snapshot at $t_{\mathrm{pb}} =  200$ ms post the
 core bounce, of a 2D simulation  for an $11.2\ \mathrm{M}_{\odot}$ progenitor 
 model~\cite{Sumiyoshi:2012za, Sumiyoshi:2014qua}. 
 The spatial and momentum resolutions of the Boltzmann calculations were 
 (256, 64, 1) and (14, 36, 12) for $(N_r , N_{\Theta}, N_{\Phi})$ and
 $(N_{E_\nu} , N_{\theta_\nu} , N_{\phi_\nu})$, respectively, where
 $(N_r , N_{\Theta}, N_{\Phi})$ are the numbers of spatial zones in the 
  spherical coordinates, and $(N_{E_\nu} , N_{\theta_\nu} , N_{\phi_\nu})$
  are the number of bins in the momentum space.
   In this snapshot,
   a total number of  $627$ ELN crossings were found in Ref.~\cite{Abbar:2018shq}
  (white crosses in Fig.~\ref{fig:2D}).

In order to test our method,  we pretended that we  have 
access to only a few moments of the neutrino angular distribution
instead of the full angular information, which was indeed provided by
the Boltzmann calculations. 
Although the inequality proposed in Ref.~\cite{Dasgupta:2018ulw}
(which corresponds to employing $\mathcal{F}(\mu) = \mu^2 \pm2\mu + 1$\ ) can not capture 
any ELN crossings in this snapshot, one can capture a number
of crossings by trying other quadratic $\mathcal{F}$'s
(using only $I_0$, $I_1$ and $I_2$). 
For instance, 
in the first panel of Fig.~\ref{fig:2D}, we show the ELN crossings which 
were captured by employing   $\mathcal{F}(\mu) = (\mu-0.88)^2$
(blue crosses). These ELN crossings, which are  
$\gtrsim 5\%$ of the total
number of ELN crossings in this snapshot,
are located 
at relatively larger radii with $r \simeq 150-200$ km.

If $I_3$ is also available,  a larger number 
of ELN crossings can be captured. For example, the ELN crossings captured by
employing $\mathcal{F}(\mu) = (1-\mu)\times(\mu-0.88)^2$ are indicted in 
 the second panel of Fig.~\ref{fig:2D}. These crossings  are 
 located at $r \simeq150-200$ km. By using $I_3$,
 we were able to capture approximately $\gtrsim 40 \%$
 of the total number of ELN crossings in this snapshot.

 From these example, one can already observe that it is crucial 
 to consider various forms of $\mathcal{F}(\mu)$ for having a good 
 chance for capturing  ELN crossings in the SN simulations. 
  Otherwise, it is completely possible to
 miss a huge fraction (if not all)  of the ELN crossings. 
 In particular,  the information available in higher moments, here $I_3$,
 is of utmost importance. 
 This piece of information is totally
 useless in the method based on the instability of conversion modes,
 as can be clearly seen in Eq.~(\ref{eq:ineq}).

\begin{figure}[tb!] 
\centering
\begin{center}
\includegraphics*[width=.49\textwidth, trim= 5 15 10 10,clip]{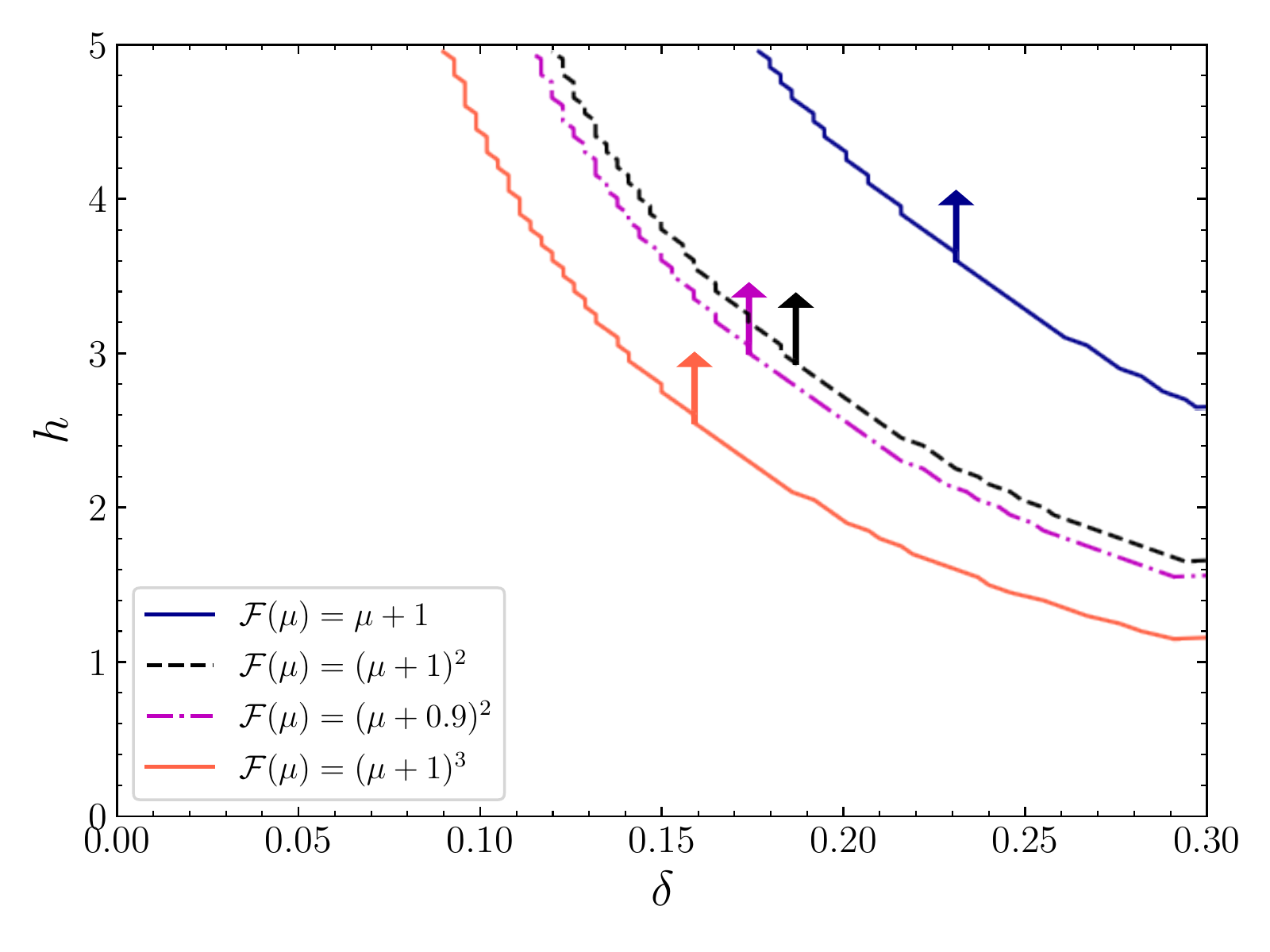}
\end{center}
\caption{
The curves above which ELN crossings can be captured for a number
of  $\mathcal{F}$'s and the ELN distribution defined in Eq.~(\ref{eq:generic}).
Narrower (smaller $\delta$'s) and shallower (smaller $h$'s)
ELN crossings  can be 
captured when higher neutrino angular moments (higher powers
of $\mu$) are employed.}
\label{fig:generic}
\end{figure}

 In principle,  the calculations that employ
 higher moments can capture larger number of ELN crossings.
 This simply comes from the fact that
if $\mathcal{F} $ includes higher powers
 of $\mu$, it can  better capture the details of the shape of ELN
 distribution. In order to have a better understanding of this, we considered
 a parametric ELN distribution of the from
   \begin{equation}\label{eq:generic}
 G(\mu) =\left\{
                \begin{array}{ll}
                  +1\quad \text{for}\quad \mu <1-\delta,\\
                   -h\quad \text{for}\quad \mu>1-\delta,
                \end{array}
              \right.
\end{equation}
to mimic the shape of realistic ELN distributions (see, e.g., Fig. 5 in Ref.~\cite{Abbar:2019zoq}). 
Such an ELN distribution has a crossing with the width of $\delta$ 
and the depth of $h$. As shown in Fig.~\ref{fig:generic},
narrower  crossings (smaller $\delta$'s) can be captured when higher
angular moments are used, provided that they are deep (large $h$) enough.
Although here the narrowest crossings can be captured  when $I_3$ is
used, calculations with only $I_0$ and $I_1$ can also capture some
ELN crossings if they are wide and deep enough.

Apart from $\mathcal{F}(\mu) $'s that are always positive in the 
interval $[-1,+1]$, one may also choose functions 
 which are positive
only within a certain range, however, provided that $G(\mu) $
is negligible within the range in which $\mathcal{F}(\mu) $ is not positive. 
In particular,  one normally expects $G(\mu) $ to be almost negligible in the 
backward direction, i.e. for $\mu < 0$,  above the neutrinosphere. 
Thus, all $\mathcal{F}(\mu) $'s which
are positive in the interval $[0,+1]$ could also be used for the SN zones
above/within the decoupling region. However, it must be kept in mind
that there is some uncertainty involved here because one needs to assume
that $G(\mu) $ is small enough in the backward direction. 
ELN crossings captured by two such functions, namely
$\mathcal{F}(\mu)  = \mu$ and $\mathcal{F}(\mu)  = \mu^3$ are shown in 
 the  right panels of Fig.~\ref{fig:2D}, which are 
 located at $r \simeq 60-90$ km. Here, we took the average radius of the
 neutrinosphere to be 60 km for this time snapshot.

The method we propose here focuses only on the occurrence of 
ELN crossings in the SN environment
and does not aim at studying the unstable modes
and their corresponding growth rates. 
Though having ELN crossing and fast modes are not  
necessarily equivalent from the mathematical point of view,
they are most likely equivalent considering the 
expectations one has for the shape of the ELN distribution. 
In particular, 
one normally expects a single crossing~\footnote{As reported in Ref.~\cite{Abbar:2019zoq},
there can exist SN zones  at larger radii for which ELN distribution has  two crossings, one at 
$\mu \simeq 1$ and one at small $\mu$'s. However, the second
crossing seems to be extremely shallow so that it only slightly modifies
the linear stability analysis (at least in the presence of another bigger ELN crossing
in the forward direction).}   
 in $G(\mu) $  which means that the
occurrence of fast modes and ELN crossings should be 
equivalent~\cite{Abbar:2017pkh, Capozzi:2019lso}.
 In addition, because only a few moments are available,
 our method does (can) not capture extremely narrow crossings.
 This implies that the consequent growth rate of the unstable modes, $\kappa$,
 should be remarkably large, i.e. $\kappa \gtrsim (0.01-0.1) \mu$~\cite{Abbar:2019zoq}. 
  
Nevertheless, it might  also be possible to make an estimate  
of the unstable modes and their growth rates by using a few
neutrino angular moments.
In particular, 
we observed that
a parametric function with the form 
 \begin{equation}\label{eq:fit}
 \tilde{f}_\nu(\mu) \propto e^ {-(1 - \mu)^\beta/\alpha},
 \end{equation}
 may appropriately capture  neutrino and antineutrino angular distributions 
 integrated over $E_\nu$ and $\phi_\nu$,
 \begin{equation}
   \tilde{f}_\nu(\mu) =
  \int_0^\infty \int_0^{2\pi}  \frac{E_\nu^2 \mathrm{d} E_\nu \mathrm{d} \phi_\nu}{(2\pi)^3}
        f_{\nu}(\mathbf{p}).
\end{equation} 

\begin{figure*}[tb!] 
\centering
\begin{center}
\includegraphics*[width=1.\textwidth, trim= 10 20 10 10,clip]{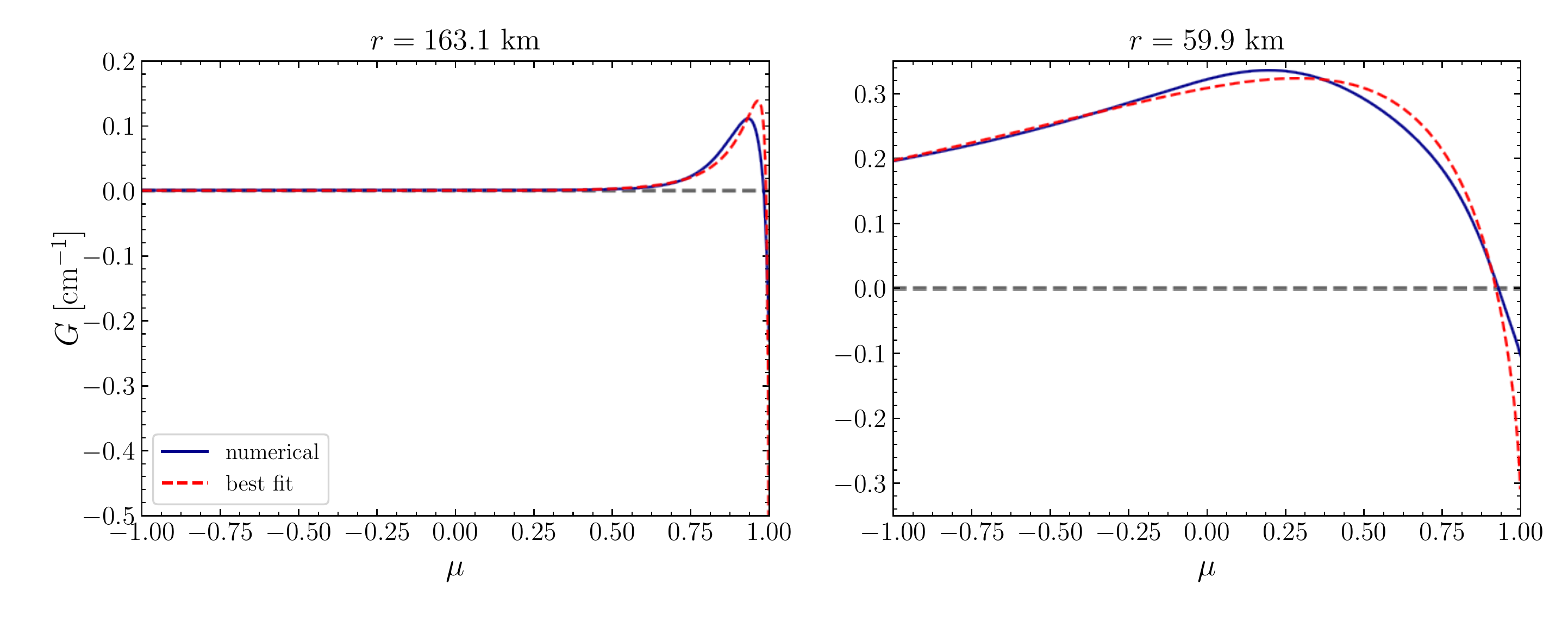}
\end{center}
\caption{
The best fits and numerical ELN angular distributions at two different radii. 
Eq.~(\ref{eq:fit}) can, indeed, provide a good estimate of the overall shape 
of the ELN  angular distributions.}
\label{fig:fit}
\end{figure*}

One can then calculate neutrino angular moments 
as a function of $\alpha$ and $\beta$ and   
 use the moments available in the SN simulation to
find the best fits for the parameters $\alpha$ and 
$\beta$ (for both $\nu_e$ and $\bar\nu_e$).
 In Fig.~\ref{fig:fit},
we use the zeroth, first and second moments of neutrino
angular distributions  
 to find $\alpha$'s and $\beta$'s for 
$\nu_e$ and $\bar\nu_e$, and we plot
the corresponding $G(\mu)$'s along with the ones obtained from the 
simulation of Ref.~\cite{Abbar:2019zoq}, at two different radii. 
Although we
leave a dedicated and systematic study of the appropriateness
of this fitting to a future work, it seems that the function proposed 
in Eq.~(\ref{eq:fit}) can be  useful in capturing 
  the overall shape of
the ELN angular distribution and its potential  crossings,
though it could have its own limitations specifically at very large 
radii where  the crossings get too narrow.

\section{Conclusion}
It is thought that neutrinos can undergo fast flavor conversions
in dense neutrino media, such as those in CCSNe, provided that the angular distributions of
the electron neutrino and antineutrino cross each other. Nevertheless, such 
detailed angular information is not available in most of the state-of-the-art 
CCSN simulations where instead, a few angular moments are considered
in the treatment of the neutrino transport. 
We have proposed a new method to search for fast modes
in such CCSN simulations. Our proposed method
 is based on the information in the available neutrino angular moments
and it can use, and work with, as many  angular 
moments as are provided by the SN simulation.

We check the practicability of our approach by employing it to some
realistic neutrino angular distributions obtained from a 2D CCSN simulation.
 In particular, 
for the snapshot at $t_{\mathrm{pb}} = 200$ ms,
we show that it can capture $\gtrsim 40 \%$ of the total number of ELN crossings
 when $I_0$, $I_1$, $I_2$ and $I_3$ are all available. 
 This should be compared with the methods based on the instability 
 of the zeroth mode~\cite{Dasgupta:2018ulw}  
 which can not capture any crossing
 in this snapshot.
 This indicates the high efficiency of our proposed method
 to extract  the most available information in the neutrino
angular moments.
 We show that this is not coincidental
 and indeed,  can be understood by considering  
 the shape of the ELN
 distributions expected in the neutrino decoupling region. 

Although the occurrence of the ELN crossings is most likely a necessary
condition  for  fast modes, 
it may not,  in general, be a sufficient condition.
However, for a physically reasonable  ELN angular distribution,
 it can be also a sufficient condition~\cite{Abbar:2017pkh, Capozzi:2019lso}. 
In addition, since  CCSN simulations  consider  only  a few
neutrino angular moments, our method is not aggressive in terms
of distinguishing very narrow  crossings. Hence, any crossings captured
by our method can be expected to result in fast flavor conversion modes 
with significant growth rates.

While we have utilized our method to capture the  crossings
in the  $\phi_\nu$-integrated ELN distribution, it can  be also used to capture
the crossings which may occur in $\phi_\nu$. To achieve this goal, one just need to modify 
Eq.~(\ref{eq:I}) to
 \begin{equation}
I_\mathcal{F} = \int_{-1}^{1}  \int_{0}^{2\pi} \mathrm{d}\mu \mathrm{d}\phi_\nu\ 
\mathcal{F}(\mu,\phi_\nu)\  \tilde{G}_{\rm{\bf{v}}},
\end{equation}
where 
\begin{equation}
  \tilde{G}_{\rm{\bf{v}}} =
  \sqrt2 G_{\mathrm{F}}
  \int_0^\infty \frac{E_\nu^2 \mathrm{d} E_\nu  }{(2\pi)^3}
        [f_{\nu_e}(\mathbf{p})- f_{\bar\nu_e}(\mathbf{p})],
 \label{Eq:G}
\end{equation}
and $\mathcal{F}(\mu,\phi_\nu) $ is a positive function of $\mu$ and $\phi_\nu$.
For example, in the $M_1$ neutrino transport scheme, one can choose
$\mathcal{F}(\mu,\phi_\nu) = \mu^2 \cos^2\phi_\nu$ and so on.

In this study, we also propose a parametric function 
that could appropriately describe
 neutrino and antineutrino angular distributions. This way, 
one can also study the unstable modes in the neutrino gas
and make an estimate of their fast  growth rates.

Fast modes can also appear in neutron star mergers (NSM)~\cite{Wu:2017qpc}
where  the neutrino gas can be extremely dense. 
The method proposed here is not limited to the case of CCSN 
simulations and can be also used in  NSM simulations
where the neutrino transport is treated by moments method~\cite{Just:2014fka}.

Despite the fact that  a number of studies have detected  fast modes in CCSN simulations, 
a comprehensive understanding of the characteristics  of  the fast modes in 3D SN models is 
still missing. 
Our study  allows for a more efficient identification of the ELN crossings 
in the SN/NSM environments and can lead to improvement of our understanding 
of the neutrino flavor evolution in such astrophysical settings.  

\acknowledgments
I am really grateful to Huaiyu Duan and Georg Raffelt
for many insightful discussions and their helpful comments on the manuscript.
I also thank  Cristina Volpe, Irene Tamborra and Francesco Capozzi for very useful
conversations. 
 I am also indebted to 
Kohsuke Sumiyoshi for
providing me with the data of the CCSN simulations used in
this study. 
I acknowledge partial support by the Deutsche Forschungsgemeinschaft
(DFG) through Grant No. SFB 1258 (Collaborative Research Center Neutrinos,
Dark Matter, Messengers).



\bibliographystyle{elsarticle-num}
\bibliography{Moment}



\end{document}